\documentstyle[12pt,psfig]{article}

\begin{document}

{\small \hfill }

{\vskip 2cm}

\begin{center}
{\Large {\bf Three-Dimensional Ordering in Weakly Coupled
Antiferromagnetic Ladders and Chains}}

{\vskip 1cm}

{\bf Stefan Wessel and Stephan Haas}

{\vskip 1cm}

{Department of Physics and Astronomy, University of Southern California}

{\ Los Angeles, CA 90089-0484, USA}

{\vskip 2cm}

{\bf }

{\vskip 0.5cm}
ABSTRACT
\end{center}
A theoretical description is presented for
low-temperature magnetic-field induced three-dimensional (3D)
ordering transitions
in strongly anisotropic quantum antiferromagnets, consisting
of weakly coupled antiferromagnetic spin-1/2
chains and ladders. First, effective 
continuum field theories are derived for the one-dimensional subsystems.
Then the Luttinger parameters, which determine the low-temperature 
susceptibilities of the chains and ladders, are calculated
from the Bethe ansatz solution for these effective models.   
The 3D ordering transition line is obtained using a random phase
approximation for the weak inter-chain (inter-ladder) coupling. 
Finally, considering a Ginzburg criterion,
 the fluctuation corrections to this approach are shown to be small.
The nature of the 3D ordered phase resembles a Bose condensate of 
integer-spin magnons. It is proposed that for systems
with higher spin degrees of freedom,
e.g. N-leg spin-1/2 ladders, multi-component condensates 
can occur at high magnetic fields.  
\newpage

\section{Introduction}

Compounds of weakly coupled spin chains typically have an ordering transition
from a high-temperature quasi-one-dimensional (1D) phase to a low-temperature
three-dimensional (3D) phase at a critical temperature which depends on the 
inter-chain coupling constant \cite{schulz96}. This transition can be suppressed
if the individual
chains are gapped spin liquids, as it is the case for
Haldane or spin-Peierls systems, or for compounds with an Ising-anisotropy
in the intra-chain exchange coupling. In these systems, the low-energy spectrum
consists of a singlet ground state with an excitation
gap to the first triplet.
This gap can be reduced and eventually overcome by turning on and
increasing an external magnetic field. Once the spin gap is destroyed, the
residual inter-chain coupling can lead to 3D ordering at low temperatures.
\cite{haas98,wessel99,giamarchi99}
In this paper, we discuss a quantitative theoretical approach to study 
such magnetic-field induced transitions, based on an exact field-theoretical
description of the low-energy intra-chain dynamics which drives the
transition, combined with a mean-field theory (including quantum fluctuation
corrections) for the inter-chain exchange. 

Our results are in good agreement with recent experiments on the compounds
$TlCuCl_3$\cite{nikuni} and $Cu_2(C_2H_{12}N_2)_2Cl_4$\cite{chaboussant},
where transition lines
$h_c(T)$ were extracted from an analysis of
the temperature-dependent magnetization and from NMR data for the $\rm 1/T_1$
relaxation.
Typically, the spin gap in most of the ladder compounds known to date
is too large to be
overcome by presently available magnetic fields.
However,
these particular materials have small spin gaps of the order 
10 - 20 Kelvin,
which makes the interesting gapless regime experimentally accessible.
It has recently been pointed out that $Cu_2(C_2H_{12}N_2)_2Cl_4$
may better be modeled
as an ensemble of weakly coupled dimers than as an antiferromagnetic
2-leg ladder \cite{broholm}.
Whatever the precise structure may turn out to be,
a magnetic-field
induced ordering transition can occur in  all
anisotropic
spin systems with a singlet-triplet excitation gap, including
weakly coupled Ising-like
chains, spin-Peierls chains, and ensembles of spin dimers.
Other possible candidate
materials with spin gaps
include $KCuCl_3$ \cite{cavadini}, $CuGeO_3$ \cite{hase},
$\alpha'-NaV_2O_5$\cite{augier},
and the homologous series of cuprates $Sr_{n}Cu_{n+1}O_{2n+1}$ \cite{azuma}.

The exact nature of these 3D ordered phases is currently under debate \cite{giamarchi99,nikuni}. In the
case of weakly coupled gapless chains which undergo a 3D ordering transition
even in the absence of a magnetic field, long-range antiferromagnetic
order is found
below the transition temperature\cite{schulz96}. For the spin-gapped compounds
$TlCuCl_3$ and $Cu_2(C_2H_{12}N_2)_2Cl_4$
the field-induced low-temperature
transition resembles that of a Bose-Einstein condensation
of integer-spin magnons. The critical exponent for the transition
line of such a condensate, i.e. $h_c \propto T^{\alpha}$
with  $\alpha = 3/2$, is rather close to 
the experimentally observed behavior\cite{giamarchi99,nikuni}. 

In this work 
a numerical solution of the Bethe ansatz is used
to determine the susceptibilities
of the one-dimensional subsystems, combined with a generalized RPA approach
for the low-temperature 3D ordering transition.
In the following section we illustrate
this approach by discussing the case of weakly coupled Heisenberg chains with
an easy-axis anisotropy (XXZ  model). Subsequently, corrections due to
fluctuation effects 
are determined, which turn out to be rather small.
Then the case of weakly coupled two-leg ladders and dimerized chains 
in a magnetic field is discussed. These results are most pertinent to 
recent and forthcoming experiments. Finally, we examine weakly coupled
N-leg ladders with N$>$2. In this case, multiple ordering transitions can 
occur which may partially overlap. In our conclusions,
we propose that these overlapping
high-field phases are 
multi-component Bose-Einstein condensates, consisting of magnons with different
integer spin.

\section{Spin-1/2 Heisenberg Chains}
Let us consider a crystal of weakly coupled 
antiferromagnetic Heisenberg chains in a magnetic field,
described by the Hamiltonian
\begin{equation}\label{heisenberg1d}
H^{1D} = \sum_i \left[ J (S^x_iS^x_{i+1} + S^y_iS^y_{i+1} 
+ \Delta S^z_iS^z_{i+1}) - h S^z_i \right] ,
\end{equation}
where $J>0$ is an antiferromagnetic exchange constant within the 
chains, $\Delta$ is an easy-axis anisotropy and h is an applied 
external magnetic field. The chains are weakly coupled by $H' = J'\sum_{<i,j>} {\bf S}_i
\cdot {\bf S}_j $ with $0 < J' \ll J$. 

The phase diagram of $H^{1D}$ can be obtained from a numerical solution 
of the Bethe ansatz equations \cite{haldane,cabra}, and
is shown in Fig. 1. At zero magnetic field the 1D subsystem is in the 
ferromagnetic Ising regime for $\Delta<-1$ . In the interval
$-1<\Delta<1$ it is in the  gapless XY regime, 
whereas for $\Delta >1$ it is in the massive Ising antiferromagnetic regime. 
The magnetization of the system becomes non-zero if
the magnetic field $h$ exceeds a minimal field given by $h_{min}=0$ 
for $-1<\Delta<1$. In the gapped case 
($\Delta=\cosh \gamma \geq 1$):
\begin{equation}
h_{min}=J\frac{2\pi\sinh\gamma}{\gamma}
\sum_{n=0}^{\infty}\frac{1}{\cosh\frac{\left(2n+1\right)\pi^2}{2\gamma}}.
\end{equation} 
It saturates at a maximum critical field given by $h_{\max}=(1+\Delta)J$.

\begin{figure}
\centerline{\psfig{figure=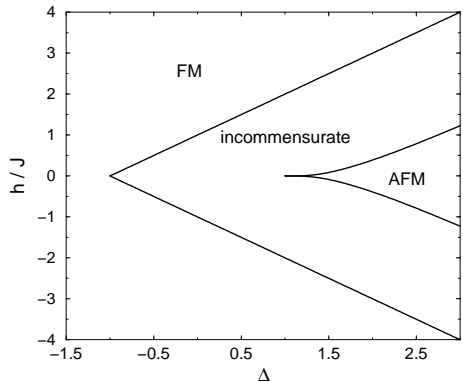,width=6cm,angle=-90}}
\caption{Phase diagram of a spin-1/2 Heisenberg chain with an
easy-axis anisotropy, $\Delta$, in a magnetic
field, $h$.
FM: ferromagnetic regime; incommensurate: partially polarized gapless
regime; AFM: antiferromagnetic
regime.
}
\end{figure}
  
The inter-chain coupling $J'$ is assumed to be small compared to the
intra-chain coupling $J$. Therefore,
3D long-range antiferromagnetic order only occurs
in the gapless region. In order to calculate the magnetic response of 
the chains and the 3D ordering temperature one needs to know the 
low-temperature behavior of the susceptibility at finite magnetic fields. 
This can be determined analytically by mapping $H^{1D}$ onto an effective 
continuum field theory that describes the low-lying excitations of $H^{1D}$ 
in the gapless incommensurate regime. For $\Delta$ close to zero the 
long-wavelength limit of $H^{1D}$ can be studied via bosonization techniques, 
leading to a $c=1$ conformal field theory (CFT). The effective parameters of 
this CFT for the {\it whole range} of $\Delta$ can be determined by comparing 
the thermodynamical properties of this CFT with the numerical 
thermodynamic Bethe ansatz solution to the original Hamiltonian
(\ref{heisenberg1d}).   

In the following, we discuss the technical aspects of this procedure.
The Hamiltonian (\ref{heisenberg1d}) can be mapped onto a model of interacting spinless fermions via a Wigner-Jordan  transformation \cite{wigner}
\begin{equation}\label{wjtrafo}
S^+_i=S^x_i+iS^y_i=a^{\dagger}_i \exp\left(i \pi \sum_{j=1}^{i-1}a^{\dagger}_j a_j\right),\quad S^z_i=a^{\dagger}_i a_i -\frac{1}{2}.
\end{equation}
Using the fact that $H^{1D}(J,\Delta,h)$ is related to 
$H^{1D}(-J,-\Delta,h)$ via a unitarity transformation, this mapping gives
\begin{eqnarray}\label{hf1d}
H^{1D}_F&=&-J\sum_{k\in \mbox{BZ}} \: a^{\dagger}_k a_k\:\cos k\: + (\Delta +\frac{h}{J})\:\:a^{\dagger}_k a_k\\ \nonumber
& & + \frac{\Delta J}{N}\sum_{k_1,...,k_4}\delta(k_1+k_3-k_2-k_4)\:e^{i(k_1-k_4)}\: a^{\dagger}_{k_1} a_{k_2} a^{\dagger}_{k_3} a_{k_4}.
\end{eqnarray}

The magnetization of the chain, $m$, is related to the spinless fermion 
density, $n$ , by $m=n-1/2$. Hence, the lower critical field $h^e_{c1}$ 
is determined by the condition that the band of spinless fermions starts 
to fill up. Since in this low-density limit the interactions between the 
fermions are negligible, one easily finds  $h^e_{c1}=-(1+\Delta)J_e$. 
From a particle-hole transformation $a^{\dagger}_k\rightarrow b_k$, 
the upper critical field $h^e_{c2}=(1+\Delta)J_e$ (high-density limit
for the spinless fermions) is obtained analogously.

In the gapless region (i.e. $h_{c1}\leq h \leq h_{c2}$) we bosonize $H^{1D}$,
and obtain a c=1 CFT of a compactified scalar,
\begin{equation}\label{effcft}
H_B=\int dx \left(\frac{\pi u K}{2}\Pi^2+\frac{u}{2\pi K}(\partial_x\phi)^2\right),
\end{equation}
where $\phi(x,t)$ is the bosonic field and $\Pi(x,t)$ is 
its conjugate momentum. The Luttinger
parameters $K$ and $u$ depend on the magnetic 
field and the exchange interaction of the original Hamiltonian,
and still need to be determined. One can
interpret $H_B$ as describing a compactified boson with 
radius
\begin{equation}
R=\frac{1}{\sqrt{4\pi K}}.
\end{equation}

We will now discuss the equations to determine $K$ and $u$, which can be
derived from the Bethe-ansatz for $H^{1D}$ (see e.g.
\cite{korepin}). One finds a system of integral equations 
\cite{cabra,qin},
that need to be solved numerically.
The dressed energy, $\epsilon_d(\eta)$, satisfies the integral equation
\begin{equation}\label{inted}
\epsilon_d(\eta)=\epsilon_0(\eta)-\frac{1}{2\pi}\int_{-\Lambda}^{\Lambda} K(\eta-\eta')\epsilon_d(\eta') d\eta',
\end{equation}
where the kernel $K(\eta)$ and the bare energy $\epsilon_0(\eta)$ 
are given in Table I \cite{tabfoot}. A similar table has been given
by Cabra {\it et al.} \cite{cabra}.

\begin{figure}[h]
\begin{center}
\begin{tabular}{||c|ccc||}\hline\hline
$\Delta$ & $K(\eta)$ & $\epsilon_0(\eta)$ & $g(\eta)$\\ \hline
$\cos\theta=\Delta<1$ &
$\frac{\tan\theta}{\tan^2\theta\cosh^2\frac{\eta}{2}+\sinh^2\frac{\eta}{2}}$ &
$\frac{h}{J}+\frac{\Delta^2-1}{\cosh{\eta}-\Delta}$ &
$\frac{\cot\frac{\theta}{2}}{\cosh^2\frac{\eta}{2}+\cot^2\frac{\theta}{2}\sinh^2\frac{\eta}{2}}$\\
$\Delta=1$ & $\frac{4}{\eta^2+4}$ & $\frac{h}{J}-\frac{2}{\eta^2+1}$ & $\frac{2}{\eta^2+1}$\\
$\cosh\gamma=\Delta>1$ &
$\frac{\tanh\gamma}{\tanh^2\gamma\cos^2\frac{\eta}{2}+\sin^2\frac{\eta}{2}}$ &
$\frac{h}{J}+\frac{\Delta^2-1}{\cos{\eta}-\Delta}$ &
$\frac{\coth\frac{\gamma}{2}}{\cos^2\frac{\eta}{2}+\coth^2\frac{\gamma}{2}\sin^2\frac{\eta}{2}}$\\ 
\hline\hline
\end{tabular}
\newline
\\
{Table I: Functions used in the integral equations for the XXZ chain.}
\end{center}
\end{figure}

The cut-off parameter $\Lambda$ is determined by the condition
\begin{equation}\label{conded}
\epsilon_d(\Lambda)=0.
\end{equation}
Once $\Lambda$ has been obtained, the dressed charge function 
$\xi(\eta)$ can be calculated from the integral equation
\begin{equation}
\xi(\eta)=1-\frac{1}{2\pi}\int_{-\Lambda}^{\Lambda} K(\eta-\eta')\xi(\eta') d\eta',
\end{equation}
directly giving the Luttinger exponent $K$:
\begin{equation}
K=\xi(\Lambda)^2.
\end{equation}
Simultaneously, the integral equations for the phase-space densities 
$\sigma(\eta)$ and $\rho(\eta)$ are solved self-consistently:
\begin{eqnarray}
\sigma(\eta)&=&g(\eta)-\frac{1}{2\pi}\int_{-\Lambda}^{\Lambda} K(\eta-\eta')\sigma(\eta') d\eta',\\
\rho(\eta)&=&\frac{1}{2\pi}\frac{d K(\eta-\Lambda)}{d\eta}-\frac{1}{2\pi}\int_{-\Lambda}^{\Lambda} K(\eta-\eta')\rho(\eta') d\eta'.
\end{eqnarray}
Then the spinon velocity $u$ and the magnetization $m$ are given by
\begin{eqnarray}
\frac{u}{J}&=&\frac{e}{2\pi \sigma(\Lambda)},\\
m&=&1-2 \int_{-\Lambda}^{\Lambda}\sigma(\eta) d\eta,
\end{eqnarray}
where
\begin{equation}
e=\frac{d\epsilon_0(\Lambda)}{d\Lambda}+\int_{-\Lambda}^{\Lambda}\epsilon_0(\eta)\rho(\eta)d\eta.
\end{equation}

In Fig. 2 the magnetization curves, $m(h)$, for an XY-like, an isotropic,
and an Ising-like 
Heisenberg chain are shown, along with the 
upper and lower bound of the spinon excitation spectrum.
The cases $\Delta = 0.5$ and $\Delta = 1.0$ are gapless, whereas for
$\Delta = 2.0$, the zero-field spin gap (Fig. 3(f))
leads to a plateau in $m(h)$
(Fig. 3(c)). Furthermore, the bandwidth of the spinon spectrum increases 
with $\Delta$, reflecting the renormalization of the
spinon velocity due to the
backscattering processes.

\begin{figure}[h]

\centerline{\psfig{figure=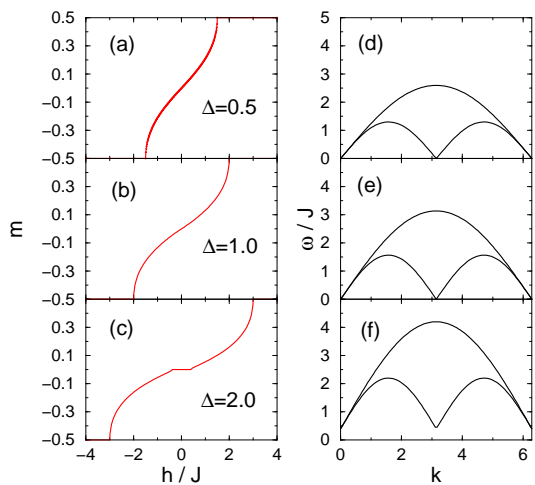,width=7cm,angle=270}}
\caption{ (a) - (c): magnetization curves, $m(h)$, 
of an antiferromagnetic spin-1/2 XXZ chain. 
(a) XY regime with Ising anisotropy $\Delta=0.5$, 
(b) isotropic Heisenberg point with $\Delta=1.0$,
 and (c) Ising regime with $\Delta=2.0$. 
The graphs in (d)-(f) show the corresponding
lower and upper bounds of the spinon excitation spectrum, $\omega(k)$.}
\end{figure}

From the exact numerical solution of the continuum model, 
the spin-spin correlation functions of the 
original model (\ref{heisenberg1d}) can be derived. In turn,
the finite-temperature susceptibilities are given via a Fourier 
transformation of the correlation functions \cite{schulz86}.
The low-temperature behavior of the susceptibility in the gapless regime 
is determined by the dominant low-frequency spinon modes at 
momentum $q_z=\pi$:
\begin{eqnarray}
\chi^{1D}_{+-}(q_{z},\omega =0; T)= 
F(\Delta)\left[\frac{\sin(\frac{\pi}{4K})}{u} 
\left(\frac{2\pi T}{u}\right)^{\frac{1}{2K}-2}\right. 
\nonumber \\
 \times \left.B^2\left( \frac{1}{8K},
1 - \frac{1}{4K} \right)-\frac{\pi}{u(1-1/4K)}\right],
\end{eqnarray}
where $B(x,y)$ is Euler's beta function,
and $F(\Delta)$ is a prefactor which strongly depends on the Ising anisotropy
\cite{lukyanov}.
\begin{eqnarray}
F(\Delta)&=&\frac{1}{2(1-\beta^2)^2}\left[\frac{\Gamma(\frac{\beta^2}{2-2\beta^2})}{\sqrt{\pi}\Gamma(\frac{1}{2-2\beta^2})}\right]^2\\
& &\exp\left\{-\int_0^{\infty}\frac{dt}{t}\left(\frac{\sinh(\beta^2t)}{\sinh t \cosh[(1-\beta^2)t]}-\beta^2 e^{-2t}\right)\right\},
\end{eqnarray}
where $\cos(\pi\beta^2)=\Delta$.
Here, it has been assumed
that the chains are parallel to the $z$-axis of
the crystal. 
Furthermore, we have neglected higher-order logarithmic corrections which 
arise
in a more rigorous treatment of the backscattering processes.

The low-temperature transition line to 3D ordering due to the small but finite
inter-chain coupling $J'$ can be calculated within an
RPA approximation. The corresponding 3D susceptibility is then given 
by \cite{schulz96,jensen}
\begin{equation}
\chi^{3D}({\bf q},\omega=0;T)=\frac{\chi^{1D}(q_z,\omega=0;T)}{1+J' f({\bf q})\chi^{1D}(q_z,\omega=0;T)},
\end{equation}
where $f({\bf q})$ is the crystal form factor which we here set 
to $f({\bf q})=-1$ for simplicity (simple cubic lattice). 
The 3D ordering transition is driven by the low-temperature divergence of
$\chi^{1D}(q_z,\omega=0;T)$, where the transition temperature is given by 
the locus of the divergence of $\chi^{3D}({\bf q},\omega=0;T)$.
The resulting magnetic field dependence of $T_c$ is shown in Fig. 3
for the various regimes of $\Delta$. The field dependence of the onset
of the crossover 
from 1D to 3D is shown by the dashed line, which is  
obtained from the fluctuation formula derived in the next section. 

\begin{figure}[h]
\centerline{\psfig{figure=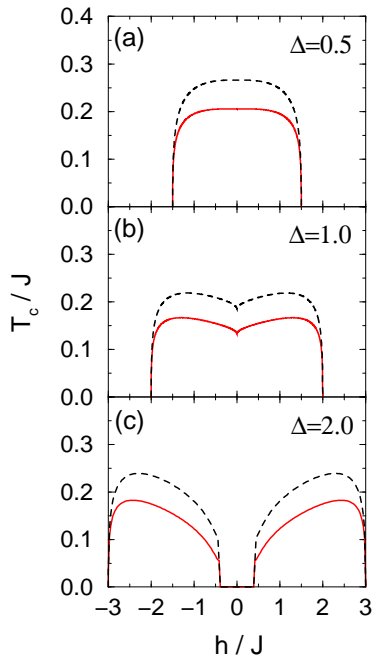,width=5cm,angle=270}}
\caption{3D ordering temperature (solid line) as a function of the 
applied magnetic field in a cubic crystal of weakly coupled antiferromagnetic 
spin-1/2 XXZ chains. (a) XY regime with $\Delta=0.5$, (b) Heisenberg point 
($\Delta=1.0$), and Ising regime with $\Delta=2.0$. The dashed lines 
indicate the onset of the fluctuation region below which the 3D magnetic 
correlation length becomes comparable to the inter-chain spacing. For this 
plot, we have chosen $J'/J=1/16$.}
\end{figure}

In this section, we have calculated the Luttinger parameters
($K$ and $u$) of the effective CFT (\ref{effcft}) for the XXZ chain
from a numerical solution of the Bethe ansatz equations within the
whole range of $\Delta$. However, the particular case of the
XY limit ($\Delta=0$) 
can be treated {\it exactly} via bosonization 
(after linearization of the dispersion law  around the Fermi points), 
since the corresponding fermionic theory is free in this case. 
In this limit, one finds 
the following values for $u$ and $K$:
\begin{equation}
\frac{u}{J}=v_F=\sqrt{1-\left(\frac{h}{J}\right)^2}\:,\quad\quad K=1.
\end{equation}

In this limit, the 3D transition temperature is
\begin{equation}
T_c=J\: v_F^{1/3}\: C_1\left[\frac{J}{J'}+C_2\frac{1}{v_F}\right]^{-2/3},
\end{equation}
where $C_1$ and $C_2$ are given numerical constants. 
At the lower critical field $h_{c1}=-1$ the behavior of $T_c$ is thus 
$T_c\propto \sqrt{h-h_{c1}}$ \cite{footnote1}.
From the numerical results a similar scaling 
behavior at the edge of the gapless phase is obtained for nonzero values 
of $\Delta$.

\section{Ginzburg Criterion}
To determine the quality of the RPA theory described in the previous section,
a Ginzburg Criterion will now be used to examine the width $\Delta T_c$ of 
the critical region. This can be done within the spinless fermion picture of 
the coupled chains, for which a Landau-Ginzburg functional can be derived. 
Denoting the localized spins by ${\bf S}_{i\mu}$, where the index $\mu$ labels
the 1D-chains and the index $i$ the position along these chains, 
the Hamiltonian for the entire 3D system is given by
\begin{equation}
H^{3D}=\sum_{\mu} H^{1D}_{\mu}(J,\Delta,h)+J'\sum_{i,<\mu,\nu>} {\bf S}_{i\mu} \cdot {\bf S}_{i \nu}
\end{equation}

Introducing spinless-fermion creation operators on each chain,  
this Hamiltonian can be mapped onto a model of weakly coupled metallic 
chains. Neglecting the inter-chain hopping (assuming an
Ising-like coupling between the chains), the resulting Hamiltonian is
\begin{eqnarray}
H^{3D}_F&=&-J\sum_{k,\mu }  \:a^{\dagger}_{k\mu} a_{k\mu}\: \cos k\: + \left(\Delta +\frac{h}{J}-\frac{J'}{J}\right) a^{\dagger}_{k\mu} a_{k\mu}\\\nonumber
& &+\frac{J \Delta}{N}\sum_{k_1,...,k_4,\mu}\: \delta(k_1+k_3-k_2-k_4)\: e^{i(k_1-k_4)}\:a^{\dagger}_{k_1\mu}a_{k_2\mu} a^{\dagger}_{k_3\mu} a_{k_4\mu}\\
& &-\frac{J'}{N}\sum_{k_1,...,k_4,<\mu,\nu>}\:  \delta(k_1+k_3-k_2-k_4)\:a^{\dagger}_{k_1\mu}a_{k_2\mu} a^{\dagger}_{k_3\nu} a_{k_4\nu}\nonumber.
\end{eqnarray}

Note that the inter-chain coupling renormalizes the bare chemical potential. 
After linearization of the dispersion around the Fermi points one finds a 
generalized  Landau-Ginzburg functional, describing the 3D ordering transition
of the original model in terms of a density-wave-type phase transition 
within the spinless fermion picture:
\begin{equation}
F[\Psi(x,y,z)]=\frac{1}{d_{\perp}^2}\int d^3\! x\: \left[ A |\Psi|^2+ B |\Psi|^4 + C_{\perp}|\: \nabla_{\perp}\Psi|^2 +C_{\parallel}\: |\nabla_{\parallel}\Psi|^2\right].
\end{equation}
Here $d_{\perp}$ denotes the interchain distance,
and the other parameters can be expressed in terms of those of the 
microscopic model. Analogous to Ref. \cite{menyhard}, they are
\begin{eqnarray}
A&=&\frac{1-J\chi^{1D}(q_z,\omega=0;T)}{\chi^{1D}(q_z,\omega=0;T)},\\
B&=&\frac{v}{2}\:\frac{7 \zeta(3)}{8\pi}\:\frac{\gamma^2}{T^2}J^3\:\chi^{1D}(q_z,\omega=0;T),\\
C_{\perp}&=&\frac{1}{2}J d_{\perp}^2,\\
C_{\parallel}&=&\frac{\gamma}{2\chi^{1D}(q_z,\omega=0;T)}\:\left(\frac{v}{\pi T}\right)^2,
\end{eqnarray}
where $v$ is the Fermi-velocity and $\gamma=2-\frac{1}{2K}$ is the scaling 
exponent of the singular part of $\chi^{1D}(q_z,\omega=0;T)$ for $T\to 0$.
If $T$ is close to $T_c$ it follows that
\begin{equation}
A=A'\left(\frac{T}{T_c}-1\right),
\end{equation}
with $A'=J\gamma$.
In the Gaussian approximation \cite{mckenzie}, the correlation length 
parallel to the chains is then given by
\begin{equation}
\xi_{\parallel}=\xi_{0\parallel}\left(\frac{T}{T_c}-1\right)^{-1/2},
\end{equation}
where the longitudinal coherence length $\xi_{0\parallel}$ is defined as
\begin{equation}
\xi_{0\parallel}=\sqrt{\frac{C_{\parallel}}{A'}}.
\end{equation}
In the direction perpendicular to the chains the correlation length is given by a similar expression, with
\begin{equation}
\xi_{0\perp}=\sqrt{\frac{C_{\perp}}{A'}}.
\end{equation}

Before discussing the Ginzburg criterion, 
let us derive the crossover condition which has already been mentioned above. 
There are different criteria for the definition of a crossover condition. 
Here the onset of the crossover is defined as the temperature, at which
the perpendicular correlation 
length equals the distance between the chains, i.e. 
\begin{equation}
\xi_{\perp}=d_{\perp}.
\end{equation}
This gives the following condition for the 1D-susceptibility
\begin{equation}
\ J' \: \chi^{1D}(q_z,\omega=0;T)=\frac{2}{3}.
\end{equation}
The width of the critical region according to the Ginzburg criterion 
\cite{ginzburg} is given by
\begin{equation}
\frac{\Delta T_c}{T_c}= \frac{1}{32(\pi\: \Delta C\: \xi_{0\perp}^2\xi_{0\parallel})^2},
\end{equation}
where 
\begin{equation}
\Delta C = \frac{(A')^2}{2d_{\perp}^2BT_c}
\end{equation}
is the specific heat jump per unit volume at the transition.
Using the above equations one finds
\begin{equation}
\frac{\Delta T_c}{T_c}\approx 0.01\:\frac{[J\chi^{1D}(q_z,\omega=0;T_c)]^3}{\gamma^2}.
\end{equation}
The width of the critical region is thus of the order of one percent.

\section{Two-leg ladders and dimerized chains}
In this section, we derive effective Hamiltonians, describing the 
low-energy spectrum of various gapped spin systems that are driven into a 
gapless phase by an external magnetic field.
First consider a general two-leg ladder system, described by the Hamiltonian:
\begin{displaymath}
H=\sum_{i, j}J_{ij}\; {\bf S}_{i,j}\cdot {\bf S}_{i+1,j}+J_{\perp}\sum_{i}{\bf S}_{i,1} \cdot {\bf S}_{i,2}+J_{12}'\sum_{i} {\bf S}_{i,1}\cdot {\bf S}_{i+1,2}+J_{21}'\sum_{i}{\bf S}_{i,2}\cdot {\bf S}_{i+1,1}-h\sum_{i,j} S^z_{i,j}
\end{displaymath}
where
\begin{equation}\label{ham2leg}
J_{ij}=\left\{
\begin{array}{c}
J_a \mbox{ , if $i$ even, $j=1$}\\
J_b \mbox{ , if $i$ odd , $j=1$}\\                             
J_c \mbox{ , if $i$ even, $j=2$}\\
J_d \mbox{ , if $i$ odd , $j=2$}\\
\end{array} 
\right.
\end{equation}
The various couplings are shown in Fig. 4. 
\begin{figure}
\centerline{\psfig{figure=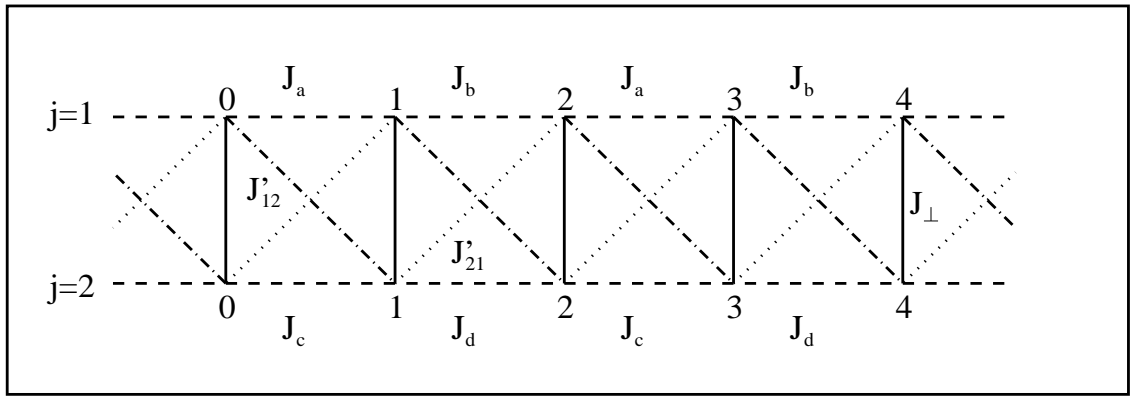,width=15cm,angle=0}}
\caption{The two-leg ladder with various couplings.}
\end{figure}
To derive the effective Hamiltonian in the gapless regime of $H$,
 a perturbation expansion in the off-rung couplings $J_{ij}$ and $J'_{ij}$ is
performed. 

Consider first the case where all couplings vanish, except for 
$J_{\perp}$. If no magnetic field is applied, the groundstate of the ladder  
consists of independent singlets at each rung. 
The Hilbert-space of a single rung is spanned by the singlet 
$\mid 0,0>=(\mid \uparrow \downarrow > - \mid \downarrow \uparrow >)/\sqrt{2}$,
and the triplet $\mid 1,1> = \mid \uparrow \uparrow >$, $\mid 1,0>=(\mid \uparrow \downarrow > + \mid \downarrow \uparrow >)/\sqrt{2}$, $ \mid 1,-1 > = \mid \downarrow \downarrow>$. 
The triplet excitation above the groundstate is  $\Delta E=1J_{\perp}$ for 
zero magnetic field. If a magnetic field is applied the triplet states split,
and for a critical field $h_{crit}=\Delta E$ the states $\mid 0,0>$ and 
$\mid 1,1>$ become degenerate. Upon increasing the magnetic field further, 
the triplet state $\mid 
 1,1>$ becomes the new groundstate of the system. 
Thus at the critical field $h_{crit}$ the magnetization changes 
discontinuously from zero to saturation.

Now consider the ladder with non-zero but small inter-rung couplings. 
The discontinuous transition for $J_{\perp} = 0$ is now 
broaded between the magnetic fields 
$h_{c1}<h_{crit}<h_{c2}$, and this gapless regime is described by an 
effective spin-1/2 Hamiltonian. To derive the corresponding
effective theory for this regime, one
introduces effective spin-1/2 operators, $\tilde{S}^{\alpha}_i$,
acting on the states 
$\mid \Uparrow > =\mid 1,1>$ and $\mid \Downarrow > = \mid 0,0 >$ 
for each rung. To first order in the small couplings, one then
finds an effective Hamiltonian,
\begin{equation}\label{hameff}
H_{eff}=\sum_{i} J_{i}^{eff}(\tilde{S}^x_{i}\tilde{S}^{x}_{i+1}+\tilde{S}^{y}_{i}\tilde{S}^y_{i+1}+\Delta^{eff}_i \tilde{S}^z_i \tilde{S}^z_{i+1})-\sum_{i}h^{eff}_i\tilde{S}^z_i,
\end{equation}
where
\begin{eqnarray}
J^{eff}_i&=&\frac{1}{2}\left({J_{i1}+J_{i2}-J'_{12}-J'_{21}}\right),\\
\Delta^{eff}_i&=&\frac{1}{2}\frac{J_{i1}+J_{i2}+J'_{12}+J'_{21}}{J_{i1}+J_{i2}-J'_{12}-J'_{21}},\\
h^{eff}_i&=&h-J_{\perp}-\frac{1}{4}\left({J_{i1}+J_{i2}+J'_{12}+J'_{21}}\right).
\end{eqnarray}
The parameters of the effective Hamiltonian become site-independent
in certain cases of interest.
For example,
if we set $J_{ij}=J_{\parallel}$, we recover the known result for the 
strongly coupled two-leg ladder \cite{mila}.
For $J_a=J_d=J_{\parallel}=(1-\delta) J$ and $J_b=J_c=0$, we obtain an 
effective description of a single chain with dimerization $\delta$ 
close to one, if we set \linebreak $J_{\perp}=(1+\delta)J$. The original model in this case is given by (see Fig. 5):
\begin{equation}
H_{D}=J\sum_{i}(1+\delta(-1)^i){\bf S}_{i}\cdot {\bf S}_{i+1}+J'\sum_{i}{\bf S}_i \cdot {\bf S}_{i+2} - h\sum_{i}S^z_i, 
\end{equation}
where $J'=J'_{12}=J'_{21}$ is a next nearest neighbor coupling constant. 
Note, that the isotropic chain ($\Delta_{eff}=1$) is recovered for 
\begin{equation}
\frac{J'}{J}=\frac{1-\delta}{6}.
\end{equation}
\begin{figure}[h]
\centerline{\psfig{figure=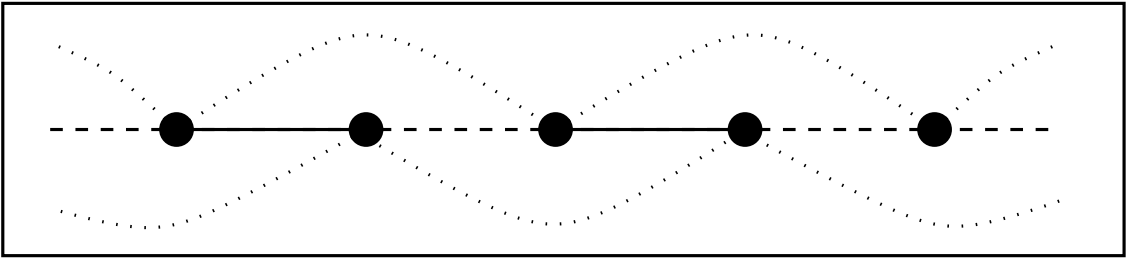,width=15cm,angle=0}}
\caption{The dimerized chain}
\end{figure}

In analogy to our discussion of the XXZ chain, the
critical fields $h_{c1}$ and $h_{c2}$ can be determined 
for the effective Hamiltonian,
 i.e. $h^{eff}_{c1}=-(1+\Delta_{eff})J_{eff}$ 
and $h^{eff}_{c2}=(1+\Delta_{eff})J_{eff}$. 
Expressed in terms of the physical variables of (\ref{ham2leg}) we 
find for the two-leg ladder:
\begin{eqnarray}
h_{c1}&=&J_{\perp}-J_{\parallel}+\frac{J'_{12}+J'_{21}}{2},\\
h_{c2}&=&J_{\perp}+2J_{\parallel},
\end{eqnarray}
and for the single dimerized chain:
\begin{eqnarray}
h_{c1}&=&\frac{1+3\delta}{2}J+J',\\
h_{c2}&=&2J.
\end{eqnarray}

Having derived these effective Hamiltonians,
the magnetic-field driven 3D ordering temperature can be studied for  
the effective models, using the results of the previous sections. 
The behavior of the transition temperature is similar to Fig. 3(c),
mainly depending on the easy-axis
anisotropy $\Delta_{eff}$ of the effective model.

\section{N-leg ladders}

Let us now consider the case of $N$-leg spin-1/2 Heisenberg ladders. 
If the coupling along the rungs is denoted as $J_{\perp}$,
and the coupling along the legs as $J_{\parallel}$, the Hamiltonian is
\begin{equation}
H^N = 
J_{\parallel} \sum_{\leftrightarrow} {\bf S}_{i,\tau} \cdot {\bf S}_{j,\tau}
    + J_{\perp}\sum_{\updownarrow} {\bf S}_{i,\tau} \cdot {\bf S}_{i,\tau '},
- h \sum_{i,\tau } 
S^z_{i,\tau },
\end{equation}
where i and j enumerate the rungs, $\rm \tau$, $\rm \tau '$
label the legs, and the sum marked by
$\leftrightarrow$ ($\updownarrow$) runs over
nearest neighbors along legs (rungs). 

We first discuss the occurance of gapless phases for these systems. Due to 
the alternating nature of the groundstates of $N$-leg ladder systems at 
zero field, which show a spin gap for even $N$ and are gapless for odd $N$, 
one has to consider these cases separately. 
In the limit of large
rung coupling ($J_{\perp}\gg J_{\parallel}$), one can consider 
first a $N$-site open Heisenberg chain in a magnetic field,
 and then treat the coupling along the leg as a perturbation.
In the case of even $N$, there will be $N/2$ changes in the nature of 
the ground state upon increasing the magnetic field. At each change a 
new gapless phase opens up. This can be seen as follows: the spectrum 
contains multiplets of multiplicities $m=1,3,5,...,N+1$. 
Let $E_m(0)$ be the energy of the lowest $m$-plet at zero magnetic field. 
Then there is a gapless phase at a field
\begin{equation}
h_i=E_{2i+1}(0)-E_{2i-1}(0),\;\; i=1,...,\frac{N}{2}
\end{equation}
In the case of odd $N$, the groundstate of the system without a magnetic field
is a doublet, and therefore the first gapless phase occurs already
at $h_1=0$. Since the spectrum contains also multiplets with $m=4,6,...,N+1$, 
there are additional gapless phases at fields
\begin{equation}
h_i=E_{2i}(0)-E_{2i-2}(0),\;\; i=1,...,\frac{N+1}{2}
\end{equation}
The values of the magnetic field where the  gapless phases occur can therefore be obtained by a numerical calculation of the zero-field spectrum. 
In each gapless phase there are two degenerate states, e.g.
\begin{eqnarray}
N=3,& h=h_2& \mid\Uparrow > = \mid \uparrow\uparrow\uparrow >\\
    &      & \mid\Downarrow > = \frac{1}{\sqrt{6}}\left(
-\mid \uparrow\downarrow\downarrow> + 2 \mid \downarrow\uparrow\downarrow > - \mid \downarrow\downarrow\uparrow>\right)
\end{eqnarray}
For non-zero values of $J_{\parallel}$, these two excitations 
spread over the ladder in the vicinity of the critical field, 
thus broadening the gapless phase.
To first order in $J_{\parallel}$, the ladder system can then be described 
by an effective spin-1/2 XXZ Heisenberg model in an effective magnetic field 
$h_{eff}$, after defining effective spin operators between the states 
$\mid \Uparrow >$ and $\mid\Downarrow >$ in the same matter as for the case of
the two-leg ladder. Considering two such $N$-site chains, coupled to each 
other by a constant $J_{\parallel}$, the energy spectrum of the corresponding 
Hamiltonian is calculated and compared to the spectrum of the 
effective model. One can thus determine the values for the parameters of the 
effective model ($J_{eff},\Delta_{eff},h_{eff}$), where the effective 
magnetic field is $h_{eff}=h-h_c(0) J_{\perp} - c_h J_{\parallel}$. 
The obtained values are given in Tab. II.
\begin{figure}[h]
\begin{center}
\begin{tabular}{||c|c||l|l|l|l||}\hline\hline
$N$ & $i$ & $J_{eff}/ J_{\perp}$ & $\Delta_{eff}$ & $h_c(0)$ & $c_h$\\ \hline
2 	& 1 	& 1 	& 0.5 	& 1 	& 0.5\\
3 	& 1     & 1	& 1	& 0     & 0  \\
3 	& 2	& 1 	& 0.5 	& 1.5	& 0.5\\
4	& 1	& 1.0750& 0.3489& 0.6589& 0.375\\
4	& 2	& 1	& 0.3750& 1.7071& 0.625\\
5	& 1	& 1.0169& 1     & 0     & 0\\
5	& 2     & 1.0961& 0.3789& 1.1189& 0.2958\\
5	& 3	& 1	& 0.3	& 1.8090& 0.7\\
6       & 1     & 1.1114& 0.3163& 0.4916& 0.3515\\
6	& 2	& 1.1348& 0.3011& 1.3860& 0.3985\\
6	& 3	& 1     & 0.25  & 1.8660& 0.75\\
7	& 1	& 1.0344& 1	& 0	& 0\\
7	& 2	& 1.1415& 0.3407& 0.8848& 0.2166\\
7	& 3	& 1.1663& 0.2381& 1.5504& 0.4891\\
7	& 4	& 1	& 0.2143& 1.9010& 0.7857\\
8	& 1	& 1.1364& 0.3020& 0.3926& 0.3432\\
8	& 2	& 1.1882& 0.2743& 1.1506& 0.2888\\
8	& 3	& 1.1917& 0.1962& 1.6577& 0.5555\\
8	& 4 	& 1	& 0.1875& 1.9239& 0.8125\\
\hline\hline
\end{tabular}
\end{center}
{Table II: Parameters of the effective low-energy model for the gapless regions of N-leg-spin-1/2 ladders in a magnetic field, h. The effective magnetic field is given by $h_{eff}=h-h_c(0)J_{\perp}-c_h J_{\parallel}$.}
\end{figure}
Using these effective model descriptions of the N-leg ladder in the gapless 
phases, we apply the RPA approach described above to a crystal of weakly 
coupled N-leg ladders. We now observe a cascade of $N/2$ ($(N+1)/2$ 3D 
ordering transitions for quasi 1-D ladder subsystems with an even (odd) number
 of legs, as shown in Fig. 6.
\begin{figure}[t]
\centerline{\psfig{figure=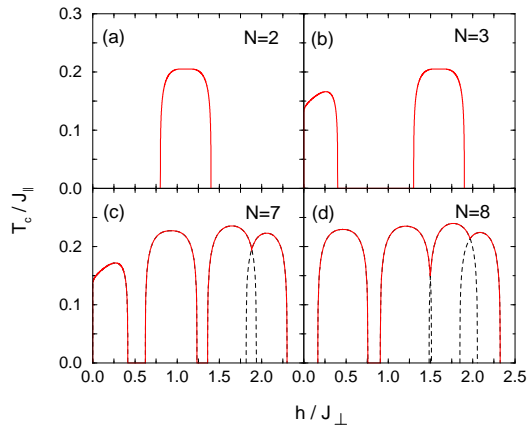,width=7cm,angle=270}}
\caption{3D ordering transition temperatures of N-leg spin-1/2 Heisenberg 
ladders as a function of an external magnetic field. Cascades of transitions 
are observed for $N>2$, driven by 1D SDW's of spin-S multiplets on the 
ladders. For this plot, we have chosen an anisotropy ratio 
$J_{\parallel}/J_{\perp}=1/5$ and residual inter-ladder coupling 
$J'/J_{\parallel}=1/16$.}
\end{figure}
 In the case of weakly coupled even-leg ladders, 
the first transition is driven by the formation of
a SDW of triplets along the ladder direction, 
with a groundstate wavevector which is proportional
to the magnetic field., $h> h_{c1}$. 
The following transition (for $N>2$) is driven by a SDW of quintuplets, etc..
Depending on the ratio $J_{\parallel}/J_{\perp}$, 
these phases of different multiplet polarization may overlap, 
and mixed regimes can occur. The resulting 3D ordering temperature does not 
vanish completely in this case, but has minima at particular magnetic fields 
where the number of the lower multiplet excitations equals the number of the 
next-higher multiplet excitations. The dependence of this overlap on the ratio 
of the coupling constants is shown in Fig. 7, where the dark stripes show 
the phases of 3D order at $T=0$ for the cases $N=3,4,7$ and $8$. 
These diagrams are in good agreement with results obtained by numerical 
finite cluster diagonalizations \cite{cabra}. 
Odd-leg ladders also have a sequence of ordering transitions, with the only 
difference that the onset of the first transition occurs already at $h=0$. 
(Fig. 6 (b) and (c)). 

\begin{figure}[h]
\centerline{\psfig{figure=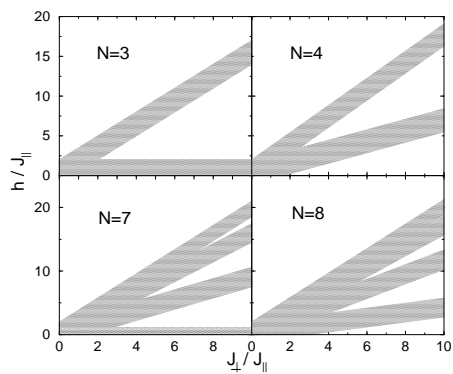,width=6cm,angle=270}}
\caption{Schematic phase diagrams of weakly coupled N-leg ladders in a 
magnetic field, $h$, at $T=0$. The shaded areas indicate 3D ordered phases,
 and shown is the dependence on the ratio of the ladder couplings 
$J_{\perp}/J_{\parallel}$.}
\end{figure}

\section{Conclusions} 

We have presented a theoretical approach to magnetic-field induced
3D ordering transitions
in strongly anisotropic antiferromagnetically correlated spin-1/2 compounds.
These systems consist of weakly coupled
chains or ladders which have singlet groundstates
with finite excitation gaps to the lowest triplet states.
Because the inter-chain
couplings are weak compared to the intra-chain couplings,
the 1D subsystems are effectively
independent. By applying an external magnetic field, the
spin gap in the subsystems can be overcome, and they become partially
polarized. In this incommensurate regime,
even infinitesimal inter-chain coupling leads to
3D long-range ordering at low temperatures.

There are only a few known realizations of Bose-Einstein condensation (BEC).
The two most prominent
examples are ultracooled dilute ensembles of trapped atoms and the superfluid
transition of Helium 4. In principle, BEC can occur
for any bosonic many-body system  in a confining potential
at low temperatures.
Therefore,
magnetic compounds with integer-spin excitations  (e.g. spin-1 magnons)
present a promising class of candidate materials. However, indications
for BEC in such materials have not been found until two recent experiments on
$\rm TlCuCl_3$ and $\rm Cu_2(C_2H_{12}N_2)_2Cl_4$. These compounds are
strongly anisotropic, and have a finite spin
gap, $\Delta$, between their singlet
groundstate and the first triplet excitation. Hence their
magnetization is exponentially activated at small temperatures and zero
magnetic field.
An applied magnetic field can decrease the singlet-triplet
excitation gaps of the 1D subsystems,
and eventually drive them
into a partially polarized,
gapless regime if the field exceeds a critical strength.
Due to residual magnetic couplings between the subsystems,
a low-temperature
1D to 3D transition occurs.
The resulting 3D ordering may be viewed as a
BEC of spin-1 magnons, accurately predicting the temperature
dependence of the critical field ($\rm h_{BE}(T) - \Delta \propto T^{2/3}$)
and the magnetization curve\cite{giamarchi99,nikuni}.

Let us now consider multi-component phases, as they occur
in spin ladders with more than two legs. In a spin ladder, a spin-1 magnon
excitation essentially corresponds to the formation of a 
spin triplet on a rung,
a spin-2 magnon corresponds to a quintuplet, etc. . At intermediate fields
(larger than $\rm h_{c1}$) these excitations can coexist, depending on the
choice of parameters. The resulting
low-temperature condensate may then contain
multiple components, as it is the case at high fields in Figs. 6(c) and (d)
(dashed regions).
These phases may be viewed as multi-component Bose-Einstein condensates,
consisting of
magnon excitations with different integer-spins. It is thus of 
interest to conduct high magnetic field experiments on appropriate 
candidate materials to determine whether such phases exist in physical
systems.

S.W. acknowledges support from DAAD under
grant number HSP III,D/98/11174, S.H. is supported by the Zumberge
Foundation.


\newpage 


\begin{thebibliography}{99}

\bibitem{schulz96} H. J. Schulz, Phys. Rev. Lett. {\bf 77}, 2790 (1996)
\bibitem{haas98}S. Haas and M. Sigrist,
{\it Physical Phenomena at High Magnetic Fields III},
edited by J.R. Schrieffer, L. Gorkov, and
Z. Fisk, World Scientific, Singapore (1999).
\bibitem{wessel99} S. Wessel and S. Haas, cond-mat/9905331.
\bibitem{giamarchi99} T. Giamarchi and A. M. Tsvelik, Phys. Rev. B {\bf 59},
11398 (1999).
\bibitem{nikuni}  T. Nikuni, M. Oshikawa, A. Oosawa, H. Tanaka,  
cond-mat/9908118.
\bibitem{chaboussant} G. Chaboussant {\it et al.}, Phys. Rev. Lett. {\bf 80},
2713 (1998); Eur. Phys. J. B {\bf 6}, 167 (1998).
\bibitem{broholm} C. Broholm, private communications.
\bibitem{cavadini} N. Cavadini {\it et al.},
Eur. Phys. J. B {\bf 7}, 519 (1999).
\bibitem{hase} M. Hase, I. Terasaki, and K. Uchiokura,
Phys. Rev. Lett. {\bf 70}, 3651 (1993).
\bibitem{augier} D. Augier {\it et al.}, Phys. Rev B {\bf 56}, R5732 (1997).
\bibitem{azuma} M. Azuma {\it et al.}, Phys. Rev. Lett. {\bf 73},
3463 (1994); Z. Hiroi, M. Azuma, M. Takano, and Y. Baudo,
J. Solid State Chem. {\bf 95}, 230 (1991).
\bibitem{haldane} F.D.M. Haldane, Phys. Rev. Lett. {\bf 45}, 1358 (1980)
\bibitem{cabra} D.C. Cabra, A. Honecker and P. Pujol, Phys. Rev. Lett. {\bf 79}, 5126 (1997); Phys. Rev. B {\bf 58}, 6241 (1998)
\bibitem{wigner} P. Jordan and E. Wigner, Z. Phys. {\bf 47}, 631 (1928)
\bibitem{korepin} V.E. Korepin, N.M. Bogoliubov, A.G. Izergin, Quantum Inverse Scattering Method and Correlation Functions, Cambridge Univ. Press (1993)
\bibitem{qin} S. Qin, M. Fabrizio, L. Yu, M. Oshikawa, I. Affleck, cond-mat/9705269
\bibitem{tabfoot} There are slight differences between our functions and those given in \cite{cabra}.
\bibitem{schulz86} H.J. Schulz, Phys. Rev. B {\bf 34} 6372 (1986)
\bibitem{lukyanov} S. Lukyanov and A.B. Zamolokchiov, Nucl. Phys. B {\bf 493}, 571 (1997)
\bibitem{jensen} J. Jensen and Allan R. Mackintosh, Rare Earth Magnetism, Oxford Univ. Press 1991
\bibitem{footnote1} A more careful treatment should take into consideration
that the dispersion at the bottom of the spinless fermion band is quadratic.
See e.g. T. Giamarchi and A. M. Tsvelik, Phys. Rev. B {\bf 59},
11398 (1999). In this case a modified exponent, $\alpha = 2/3$, is obtained.
\bibitem{menyhard} N. Menyhard, J. Phys. C {\bf 11}, 2207 (1978)
\bibitem{mckenzie} R.H. McKenzie, Phys. Rev. B {\bf 52}, 16428 (1995)
\bibitem{ginzburg} V. L. Ginzburg, Sov. Phys. Solid State {\bf 2}, 1824 (1960)
\bibitem{mila} F. Mila, Eur. Phys. J. B {\bf 6}, 210  (1998)
\end{thebibliography}
\end{document}